\documentclass[runningheads]{svmult}

\usepackage{makeidx}   
\usepackage{graphicx}  
\usepackage{subeqnar}  
\usepackage{multicol}  
\usepackage{physprbb}  
\makeindex             

\newcommand{\kms}{\mbox{km s$^{-1}$}}
\newcommand{\etal}{\mbox{et al.}}

\newcommand{\meth}{\mbox{CH$_{3}$OH}}

\newcommand{\Msun}{\mbox{$M_{\odot}$}}
\newcommand{\Lsun}{\mbox{$L_{\odot}$}}
 
\newcommand{\jj}[2]{\mbox{$J = #1 \rightarrow #2$}}

\newcommand{\uco}[1]{\mbox{$^{#1}$CO}} 
\newcommand{\cuo}[1]{\mbox{C$^{#1}$O}} 

\newcommand{\skipthis}[1]{}
\newcommand{\vlsr}{\mbox{$V_{lsr}$}}

\newcommand{\cm}[1]{\mbox{cm$^{#1}$}}   

\newcommand{\hh}{\mbox{H$_2$}}
\newcommand{\kp}{\mbox{K$'$}}
 
\newcommand{\degr}{\mbox{$^\circ$}}
\newcommand{\micron}{\mbox{$\mu$m}} 

\begin{document}

\title*{The Spectacular BHR~71 Outflow}
  
\author{Tyler L. Bourke}
\institute{Harvard-Smithsonian Center for Astrophysics, 60 Garden Street MS 42, 
Cambridge MA 02138, USA. tbourke@cfa.harvard.edu}

\maketitle

\begin{abstract}

BHR~71 is a well isolated Bok globule located at
$\sim$200 pc, which harbours a highly collimated bipolar outflow.
The outflow is driven by a very
young Class 0 protostar with a luminosity of $\sim$9 \Lsun.
It is one of a very small number that show enhanced abundances of a
number of molecular species, notably SiO and \meth, due to shock
processing of the ambient medium.  In this paper the properties of the
globule and outflow are discussed.
\vspace*{-12pt}
\begin{center}
``In the darkness, there'll be hidden worlds that shine''\\
-- Bruce Springsteen, Candy's Room 1977
\end{center}

\end{abstract}


\section{Introduction}

In 1977 Arge Sandqvist published a catalogue of southern ``dark dust clouds 
of high 
visual opacity'' (Sanqvist 1977 -- 95 entries, numbered 101-195), an extention 
of an earlier paper with Lindroos (Sandqvist \& Lindroos 1976) in which they 
presented H$_2$CO absorption line studies of
42 dust clouds (\#1-42).  Number 136 on Sandqvist's list (Sa~136) is a very
opaque Bok globule located near the Coalsack, later catalogued as
DC~297.7-2.8 by Hartley et al. (1986) and as entry 71 in the
globule list of Bourke, Hyland \& Robinson (1995a -- BHR~71).
Mark McCaughrean in his opening address at this conference highlighted a
number of important events that occurred in 1977, in particular IAU
Symposium 75 on Star Formation whose proceedings appeared that year.
It is fitting that BHR~71, which is featured in a beautiful VLT optical
image in the frontpiece (\& poster) of these proceedings, 
can trace its origins in the literature to that same year.

\section{Globule properties}

The globule properties have been determined by Bourke \etal\ (1995b,
1997).  Spatially and kinematically BHR~71 is associated with the
Coalsack at an assumed distance of 200~pc, 
though it may be as close as 150$\pm$30~pc (Corradi
\etal\ 1997).  Large scale \uco{12}\ \& \uco{13}\ maps of the globule
give a size of $\sim$0.5~pc and mass 40\Msun, while \cuo{18}\ observations 
which trace
high column density gas imply a size $\sim0.3 \times 0.15$~pc and a mass
of 12\Msun.  The high density (n$>10^4$ \cm{-3}) core traced in
ammonia is $\sim0.2 \times 0.1$ pc in size with a mass of 3\Msun.
The globule velocity is \vlsr\ $\sim$ $-4.5$\kms.

\begin{figure}[t]
\begin{center}
\includegraphics[width=0.8\textwidth]{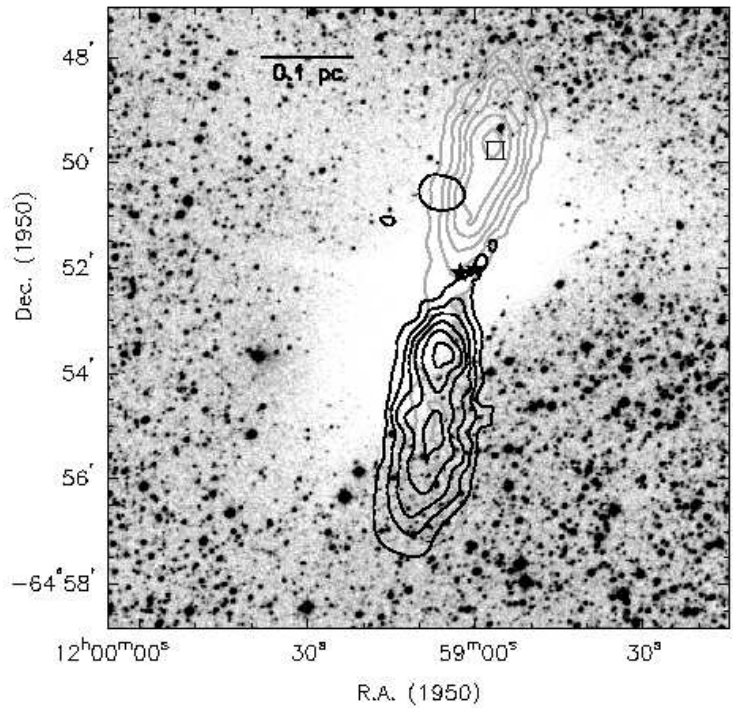}
\end{center}
\caption[]{Digital Sky Survey R-Band image of BHR~71, overlayed with
contours of \uco{12}\ \jj{1}{0}\ emission.  The black contours are
blue-shifted emission, and the grey contours are red-shifted emission.
The two ISO mid-infrared sources are indicated with star symbols.
IRS~1, to the east, is the driving source of the large outflow.  The
square marks the position of the red outflow spectra shown in 
Fig.~\ref{fig-chem}}.
\label{fig-overview}
\end{figure}

\section{CO Outflow Properties}

The properties of the large scale molecular outflow have been determined
by B97.  As can be seen in Fig.~\ref{fig-overview}, the outflow lobes are well
separated on the sky, and extend $\sim$0.3pc from their origin with an
opening angle of $\sim$15\degr.  B97 find that the velocity structure is
consistent with a steady flow with constant velocity (Cabrit \etal\ 1988).  
With this assumption the inclination of the
outflow to the line-of-sight is determined to be 85\degr.  The CO excitation
temperature in the line core is greatest at the outflow peaks,
indicating that the ambient gas there has been heated by interactions
with the outflow.  

Correcting for inclination, optical depth, and emission hidden within
the line core, B97 determine the mass in the
lobes to be $\sim$1.0\Msun\ (red lobe) and $\sim$0.3\Msun\ (blue lobe).
Considering the different methods used to determine the outflow
momentum $P$, kinetic energy $E_k$, and mechanical luminosity 
$L_{\rm mech}$ (upper and lower limit methods) B97 find 
$P=11$\Msun \kms, $E_k=60$\Msun
km$^2$s$^{-2}$, and $L_{\rm mech}=0.5$\Lsun.  There is less mass in
the blue lobe, which may be a result of this lobe breaking out of the
globule, indicated in Fig.~\ref{fig-overview} by the conical reflection 
nebulosity just south of the protostars (see the beautiful colour VLT in the
frontpiece of these proceedings for a more detailed view).

\section{Two protostars - two outflows}

\begin{figure}
\begin{center}
\includegraphics[width=0.6\textwidth,angle=-90]{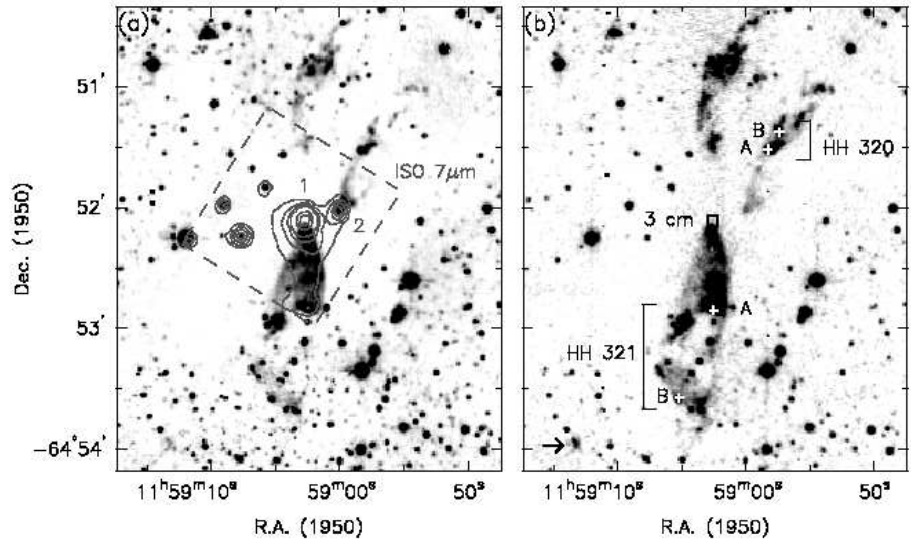}
\end{center}
\caption[]
{(a) -- \kp\ image of BHR~71 (greyscale) overlayed with ISO LW2
contours (5.0--8.5 \micron).  
The embedded protostars IRS~1 (``1'') and IRS~2 (``2'')
are labelled.  (b) -- Narrowband 2.12\micron\ + continuum image
(greyscale).  The positions of HH~320 and HH~321 
are marked with crosses, and the position of the 3~cm 
continuum source is marked with an unfilled box.}
\label{fig-nir}
\end{figure}

Near-infrared (NIR) images from the AAT are shown in 
Figure~\ref{fig-nir} (Bourke 2001).  Most of the 
non-stellar emission is due to the emission in the \hh\ $v$=1-0S(1) line, 
most likely due to shocks in the outflowing gas (Eisl\"{o}ffel 1997).  
The NIR emission is well aligned with the large scale CO outflow
(Fig.~\ref{fig-sio-co-h2}).

Mid-infrared (MIR) emission in the ISO LW2 band is overlayed on 
Fig.~\ref{fig-nir}(a).  Two of the
7\micron\ sources appear to be located at the apexes of NIR emission,
strongly suggesting that they are associated with the emission.  Source
``1" (hereafter IRS~1) lies at the apex of the reflection nebulosity seen
also in Fig.~\ref{fig-overview}
and is co-incident with the position of the mm source BHR~71-mm, 
also known as IRAS 11590-6452 (B97).  The 7\micron\ flux from IRS~1 is
an order-of-magnitude greater than from IRS~2.
The NIR feature coincident with IRS~2 in Fig.~\ref{fig-nir}(a) is
non-stellar, by comparison of its PSF with stars in the same image.

A cm continuum source (indicated on Fig.~\ref{fig-nir}(b)) 
is detected toward BHR~71 IRS~1, at both 3 and 6~cm (Wilner \etal\ 2001,
in prep).  
The spectral index is consistent with a flat or rising spectrum due 
to free-free emission, a signpost of protostellar origin (Rodr\'{\i}guez 1994). 
Corporon \& Reipurth (1997) discovered two Herbig-Haro associations in
BHR~71 -- HH~320 and HH~321,  and their locations are shown on
Fig.~\ref{fig-nir}(b).  It can be seen that
HH~320 (HH~321) is coincident with the NIR emission associated with IRS~2
(IRS~1). 

Bourke (2001) has shown that IRS~2 also drives a CO outflow which is
more compact and much less energetic than the IRS~1 outflow.
The northern part of the IRS~2 outflow is blue-shifted (and associated
with HH~320) which is the opposite of the IRS~1 outflow and allows them
to be separated spatially.   The red lobe is confused by the IRS~1
outflow, though it is probably seen in the NIR (arrowed emission in 
Fig.~\ref{fig-nir}(b)).  Bourke (2001) suggested that IRS~1 \& 2 may
form a binary protostellar pair (separation $\sim$3400AU) though the
kinematic evidence for or against is lacking.

\section{Outflow Chemistry}

\begin{figure}
\begin{center}
\includegraphics[width=0.85\textwidth]{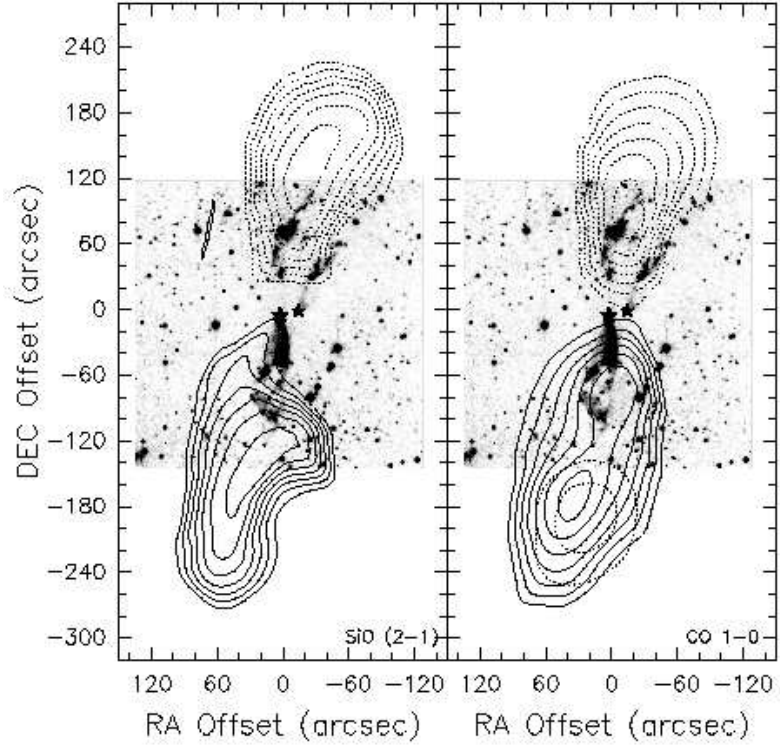}
\end{center}
\caption[]{SiO and CO integrated emission over the blue (solid lines)
and red (dotted lines) lobes of the BHR~71 outflow, overlayed on the
\hh\ image from Fig.~\ref{fig-nir}.  The protostars IRS~1 (large star)
and IRS~2 (small star) are indicated.}
\label{fig-sio-co-h2}
\end{figure}

\begin{figure}
\begin{center}
\includegraphics[width=0.9\textwidth]{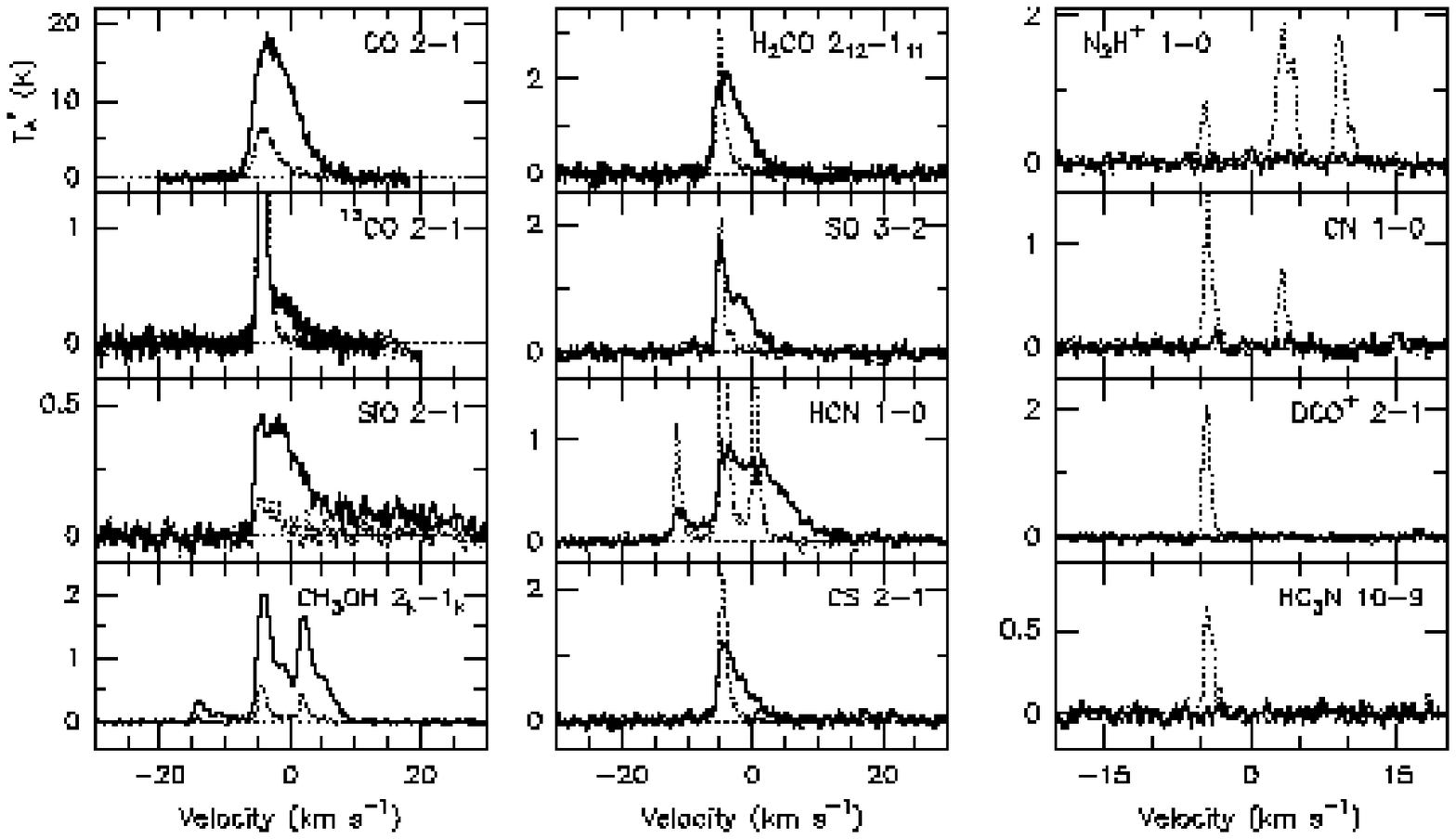}
\end{center}
\caption[]{Line profiles observed toward the red shifted lobe
(continuous lines - indicated by the square on Fig.~\ref{fig-overview})
and IRS~1 (dotted lines) of the BHR~71 outflow.}
\label{fig-chem}
\end{figure}

The BHR~71 IRS~1 outflow is one of a handful that show significant
abundance enhancements in molecules such as SiO and \meth\ (G98).
Figure~\ref{fig-sio-co-h2} shows the spatial distribution of SiO and CO in
the outflow, compared to the NIR \hh\ emission (the CO data is of lower
spatial sampling than Fig.~\ref{fig-overview}).  Figure~\ref{fig-chem}
shows spectra at two locations, the red lobe (as indicated by the box
in Fig.~\ref{fig-overview}) and at the position of IRS~1.

The spectral line profiles in the outflow and the velocity of the
outflowing gas ($<30$\kms) indicate that C-shocks dominate the flow
(G98).  The shocks are sufficiently strong to release molecules and
atoms into the gas phase via evaporation of icy grain mantles (e.g.,
\meth) and sputtering of grain cores or grain-grain collisions 
(e.g., Si, which rapidly forms SiO).  Other molecules
detected in the outflow include CS, H$_2$CO, SO, HCN, HNC, HCO$^+$ with
SEST and H$_2$O with SWAS (Bourke \etal\ 2001, in prep).  SiO is removed from 
the gas phase in about 10$^4$ years indicating that the outflow
is quite young.

In the red lobe G98 determined abundance enhancements of $\sim$350 in
SiO and $\sim$40 in \meth.  One particularly striking feature is the
spatial distribution of SiO compared to CO in the outflow.  Because SiO
is the result of Si liberation it is usually only detected at the
ends of outflows (where the shock interaction is greatest) or as a
narrow jet along the outflow axis possibly due to interactions in a
turbulent boundary layer (Garay 2000).  The wide-spread
distribution of SiO in the BHR~71 outflow is unique.  This suggests that
the SiO enhancement takes place in a shell-like structure produced by
the dynamical interaction between the ambient cloud and an underlying
wide-angle wind or wind driven shell (Garay 2000), or perhaps by a
wandering jet.  However, it has not been shown that
an interaction between a wind and the ambient material can
produce sufficient Si for this to be a viable explanation. 
If sufficient Si-bearing species are present in grain mantles then the
wind model becomes attractive (Schilke \etal\ 1997).

\section{The VLT Image}

An optical composite image taken with the VLT is shown in the frontpiece
of these proceedings.   This image hints at the
spectacular results we can expect from the VLT in the coming years.  
There is evidence in this image of both a wind component and a jet
component to the IRS~1 outflow in the blue lobe.  Extending from the 
reflection nebulosity which protrudes from the globule, one can trace out 
an elongated bubble, with its edges defined by enhanced extinction.
This is characteristic of a wide-angle wind component.  In addition,
enhanced extinction is also seen along the axis of the bubble and 
extending beyond its southern tip.  This may be an indication of the
underlying jet which is probably driving this young outflow.  Modelling
of this one image may help answer some of the remaining questions about
the spectacular BHR~71 outflow.

I thank my many collaborators on this project, in particular Guido
Garay.  A big hug to Jo\~ao for letting me present my work on this
beautiful object.


\end{document}